\documentclass[pra,aps,nopacs,showkeys,superscriptaddress,twocolumn,twoside]{revtex4}

\usepackage{graphicx,epic,eepic,epsfig,amsmath,latexsym,amssymb,verbatim}

\usepackage{theorem}
\newtheorem{definition}{Definition}

\newtheorem{lemma}[definition]{Lemma}

\newtheorem{theorem}[definition]{Theorem}

\newtheorem{conjecture}[definition]{Conjecture}

\newtheorem{expl}[definition]{Example}

\def\squareforqed{\hbox{\rlap{$\sqcap$}$\sqcup$}}
\def\qed{\ifmmode\squareforqed\else{\unskip\nobreak\hfil
\penalty50\hskip1em\null\nobreak\hfil\squareforqed
\parfillskip=0pt\finalhyphendemerits=0\endgraf}\fi}
\def\endenv{\ifmmode\;\else{\unskip\nobreak\hfil
\penalty50\hskip1em\null\nobreak\hfil\;
\parfillskip=0pt\finalhyphendemerits=0\endgraf}\fi}

\newenvironment{remark}{\noindent \textbf{{Remark~}}}{\qed}

\newlength{\blank}
\newlength{\equalsign}
\newenvironment{beweis}[1][{\hspace{-\blank}}]{{\noindent\emph{Proof~{#1}.\ }}}{\hfill\qed\vskip 0.5\baselineskip}
\newenvironment{example}[1][{}]{\begin{expl}[#1]\normalfont}{\end{expl}}

\mathchardef\ordinarycolon\mathcode`\:
\mathcode`\:=\string"8000
\def\vcentcolon{\mathrel{\mathop\ordinarycolon}}
\begingroup \catcode`\:=\active
  \lowercase{\endgroup
  \let :\vcentcolon
  }

\newcommand{\nc}{\newcommand}
\nc{\rnc}{\renewcommand}
\nc{\beq}{\begin{equation}}
\nc{\eeq}{{\end{equation}}}
\nc{\beqa}{\begin{eqnarray}}
\nc{\eeqa}{\end{eqnarray}}
\nc{\lbar}[1]{\overline{#1}}
\nc{\bra}[1]{\langle#1|}
\nc{\ket}[1]{|#1\rangle}
\nc{\ketbra}[2]{|#1\rangle\!\langle#2|}
\nc{\braket}[2]{\langle#1|#2\rangle}
\nc{\proj}[1]{| #1\rangle\!\langle #1 |}
\nc{\avg}[1]{\langle#1\rangle}
\rnc{\max}{\operatorname{max}}
\nc{\Rank}{\operatorname{Rank}}
\nc{\smfrac}[2]{\mbox{$\frac{#1}{#2}$}}
\nc{\tr}{\operatorname{Tr}}
\nc{\ox}{\otimes}
\nc{\dg}{\dagger}
\nc{\dn}{\downarrow}
\nc{\cA}{{\cal A}}
\nc{\cB}{{\cal B}}
\nc{\cC}{{\cal C}}
\nc{\cD}{{\cal D}}
\nc{\cE}{{\cal E}}
\nc{\cF}{{\cal F}}
\nc{\cG}{{\cal G}}
\nc{\cH}{{\cal H}}
\nc{\cI}{{\cal I}}
\nc{\cJ}{{\cal J}}
\nc{\cK}{{\cal K}}
\nc{\cL}{{\cal L}}
\nc{\cM}{{\cal M}}
\nc{\cO}{{\cal O}}
\nc{\cP}{{\cal P}}
\nc{\cR}{{\cal R}}
\nc{\cS}{{\cal S}}
\nc{\cT}{{\cal T}}
\nc{\cX}{{\cal X}}
\nc{\cZ}{{\cal Z}}
\nc{\csupp}{{\operatorname{csupp}}}
\nc{\qsupp}{{\operatorname{qsupp}}}
\nc{\var}{\operatorname{var}}
\nc{\rar}{\rightarrow}
\nc{\lrar}{\longrightarrow}
\nc{\polylog}{\operatorname{polylog}}
\nc{\1}{{\openone}}
\nc{\GHZ}{{\Gamma}}
\nc{\EPR}{{\Phi_2}}

\def\d{\delta}
\def\e{\epsilon}

\nc{\RR}{{{\mathbb R}}}
\nc{\CC}{{{\mathbb C}}}
\nc{\FF}{{{\mathbb F}}}
\nc{\NN}{{{\mathbb N}}}
\nc{\ZZ}{{{\mathbb Z}}}
\nc{\PP}{{{\mathbb P}}}
\nc{\QQ}{{{\mathbb Q}}}
\nc{\UU}{{{\mathbb U}}}
\nc{\EE}{{{\mathbb E}}}
\nc{\id}{{\operatorname{id}}}
\nc{\Span}{{\operatorname{span}}}

\nc{\be}{\begin{equation}}
\nc{\ee}{{\end{equation}}}
\nc{\bea}{\begin{eqnarray}}
\nc{\eea}{\end{eqnarray}}
\nc{\<}{\langle}
\rnc{\>}{\rangle}
\nc{\Hom}[2]{\mbox{Hom}(\CC^{#1},\CC^{#2})}
\nc{\rU}{\mbox{U}}

\nc{\ob}[1]{#1}

\begin{document}

\title{Entanglement of assistance\protect\\
       and multipartite state distillation}

\author{John A. Smolin}
\affiliation{IBM T.~J.~Watson Research Center, Yorktown Heights, NY 10598, USA}
\email{smolin@watson.ibm.com}

\author{Frank Verstraete}
\affiliation{Institute for Quantum Information, Caltech 107-81,
             Pasadena, CA 91125, USA}
\affiliation{Max-Planck-Institut f\"ur Quantenoptik, Hans-Kopfermann-Str.1,
             85748 Garching, Germany}
\email{fverstraete@ist.caltech.edu}

\author{Andreas Winter}
\affiliation{Department of Mathematics, University of Bristol,
             University Walk, Bristol BS8 1TW, U.K.}
\email{a.j.winter@bris.ac.uk}

\date{5th May 2005}

\begin{abstract}
  We find that the asymptotic entanglement of assistance of a general
  bipartite mixed state is equal to the smaller of its two local
  entropies.
  Our protocol gives rise to the asymptotically optimal EPR pair
  distillation procedure for a given tripartite pure state,
  and we show that it
  actually yields EPR and GHZ states; in fact, under a restricted
  class of protocols, which we call ``one-way broadcasting'', the
  GHZ-rate is shown to be optimal.

  This result implies a capacity theorem for quantum channels
  where the environment helps transmission by broadcasting the outcome
  of an optimally chosen measurement. We discuss
  generalisations to $m$ parties, and show (for $m=4$)
  that the maximal amount
  of entanglement that can be localised between two parties is
  given by the smallest entropy of a group
  of parties of which the one party is a member, but not the
  other. This gives an explicit expression for the asymptotic
  localisable entanglement, and shows that any nontrivial ground
  state of a spin system can be used as a perfect quantum repeater
  if many copies are available in parallel.

  Finally, we provide evidence that any unital channel
  is asymptotically equivalent to a mixture of unitaries,
  and any general channel to a mixture of partial isometries.
\end{abstract}


\maketitle

\section{Multipartite quantum states}
\label{sec:intro} One of the great ongoing programmes of quantum
information theory is the classification of multipartite (pure)
quantum states $\psi^{AB\ldots Z}$, and the understanding of the
possible transformations between them allowing only local
operations and classical communication (LOCC). As entanglement
presents a resource that can be used for e.g. quantum
communication, it is especially interesting to study the
asymptotic Shannon-theoretic limit. In this scenario, a few
parties hold asymptotically many copies of identical states
distributed among them, only joint operations between the
particles at the same site and classical communication between the
parties are allowed, and the conversion of states occurs with
vanishing errors in the asymptotic limit.

In the bipartite case, this question is well understood: every
pure state $\psi^{AB} = \proj{\psi}^{AB}$ is asymptotically
reversibly equivalent to maximally entangled (EPR) states,
$$\ket{\EPR}^{AB} = \frac{1}{\sqrt{2}}\bigl( \ket{0}^A\ket{0}^B
                                             +\ket{1}^A\ket{1}^B \bigr),$$
at rate $E(\psi)=S(\psi^A)=S(A)$, the entropy of
entanglement~\cite{BBPS} (Note our notation
convention: $\psi^A = \tr_B\psi^{AB}$ is the restriction of the
state $\psi^{AB}$ to $A$.) So, not only can we quantify the exact
yield of the useful EPR states, but the latter serve as a normal
form in general.

For multipartite states, the situation becomes more complex:
there does not exist any longer a single state suited as a ``gold standard'',
e.g. it is quite evident that an EPR state $\ket{\EPR}^{AB}$
can never be equivalent to any quantity of Greenberger-Horne-Zeilinger
(GHZ) states of three parties,
$$\ket{\GHZ}^{ABC} = \frac{1}{\sqrt{2}}\bigl( \ket{0}^A\ket{0}^B\ket{0}^C
                                             +\ket{1}^A\ket{1}^B\ket{1}^C \bigr).$$
So one has to aim at a ``(minimal) reversible entanglement
generating set'' (MREGS)~\cite{BPRST}, about which little is known,
except that apart from the easy candidates of $\ket{\EPR}^{AB}$,
$\ket{\EPR}^{BC}$, $\ket{\EPR}^{AC}$ and
$\ket{\GHZ}^{ABC}$~\cite{LPSW}, an MREGS has to contain at least
another state, and possibly infinitely many. See~\cite{bristolians} for an 
instructive case study.

Usually the two parts of the multi-party entanglement
programme --- classification and possible transformations --- are
viewed as one, but as we have seen, the first is really much harder:
this is because it involves studying
the transformations between pairs of states which are
\emph{asymptotically reversible}.

In this paper, we have a more modest goal: we want to go from
(many copies of) a given state to particular, interesting states,
like the EPR and GHZ state. To be precise, one would like to
``distill'' as many as possible of these target states, with high
fidelity in the limit of $n\rar\infty$, and will care primarily
for optimality of these processes and not so much for
reversibility.

\section{The task(s)}
\label{sec:task}
Given many copies of a (pure) tripartite state
$\psi^{ABC}$, which ``standard'' entangled
states like EPR states $\EPR$ between any pair of parties, or GHZ states
$\GHZ$ can the three parties distill by local operations and
classical communication (LOCC)?

We shall focus on three scenarios in succession: first, we study
optimal distillation of EPR states between a given pair of players
from a tripartite state; second we recast the protocol as one of
distilling EPR and GHZ states at the same time, and show that for
this target set, and a restricted class of protocols it gives
optimal yield; and third, we look at $m$-partite states and how
many EPR states between a prescribed pair of players can be
distilled by local measurements on the other $m-2$ parties.

Scenario 1 was studied in~\cite{E-of-A,cohen:EoA,LVvE} under the
name of ``entanglement of assistance'' in a non-asymptotic setting:
given a mixed state
$\psi^{AB}=\tr_C\bigl(|\psi^{ABC}\rangle\langle\psi^{ABC}|\bigr)$
shared between $A$ and $B$, there exists a unique purification
$\psi^{ABC}$ up to local unitary operations on $C$, and the
question was asked how much EPR-type entanglement can be created
between $A$ and $B$ when $C$ is doing local measurements and
communicates the results to $A$ and $B$. In this paper we
completely solve that question in the asymptotic setting.

About scenario 2 very little has appeared in the literature,
except for upper capacity bounds, e.g.~\cite{LPSW}, and a few
(qubit) protocols which however remain largely in the single-copy
setting~\cite{cohen:brun,acin:etal}.

The third scenario has been studied in the context of spin chains
under the name of \emph{localisable entanglement}~\cite{localiz}.
The present work will reveal some intriguing connections between
the concept of entropy of a block of spins and the entanglement
length in spin systems.

The main results are as follows:
\begin{theorem}
  \label{thm:EoA:infty}
  Given a pure tripartite state $\psi^{ABC}$, then the optimal EPR rate distillable
  between $A$ and $B$ with the help of $C$ under LOCC is
  $$E_A^\infty\bigl( \psi^{ABC} \bigr)
                      = \min\bigl\{ S(A),S(B) \bigr\}.$$
  (Our notation is such that the first two parties obtain EPR states,
  and the remaining is the helper.)
  This is 
  the asymptotic entanglement of assistance~\cite{E-of-A}.
\end{theorem}

Writing the tripartite state as
$\ket{\psi}^{ABC} = \sum_j \sqrt{q_j}\ket{\psi_j}^{AB}\ket{j}^C$,
with orthogonal $\ket{j}$ --- corresponding to to a pure state
decomposition $\psi^{AB} = \sum_j q_j \proj{\psi_j}^{AB}$ ---
let $\overline{E} = \sum_j q_j E(\psi_j)$ be the average entanglement
of the pure state decomposition. Define
$\chi = \min\{S(A),S(B)\} - \overline{E}$.

\begin{theorem}
  \label{thm:EPR+GHZ}
  Let $\psi^{ABC}$ be a pure tripartite state.
  Then, for $\e,\d>0$ and sufficiently large $n$, there exists a protocol
  involving only an instrument on $C^n$ and broadcast of the measured
  result, followed by local operations on $A^n$ and $B^n$, which
  effects the transformation
  $$\left(\psi^{ABC}\right)^{\ox n} \longrightarrow
               \left(\GHZ^{ABC}\right)^{\ox n(\chi-\d)}
                 \ox
               \left(\EPR^{AB}\right)^{\ox n(\overline{E}-\d)}$$
  with fidelity $1-\e$.
\end{theorem}

Observe that the minimal value of $\overline{E}$
above is the \emph{entanglement of formation} $E_F(\psi^{AB})$
of the mixed state $\psi^{AB}$ between Alice and Bob~\cite{BDSW}.
In the limit of many copies we have to substitute
the \emph{entanglement cost} $E_C(\psi^{AB})$~\cite{HHT}.
This outcome is better than theorem~\ref{thm:EoA:infty},
as we can always (irreversibly) turn GHZ states into
EPR states, and achieve the previous EPR-rate.
Theorem~\ref{thm:oneway-bc} in section~\ref{sec:EPR+GHZ:protocol}
shows that the corresponding GHZ-rate, $\min\{S(A),S(B)\}-E_C(\psi^{AB})$,
is indeed optimal under an important class of protocols.

\begin{theorem}
  \label{thm:n-party:EoA}
  For an $m$-party state $\psi^{ABC_1\ldots C_{m-2}}$, the optimal
  rate $R$ of EPR states distillable between $A$ and $B$
  with the help of the $C_i$ via LOCC, satisfies
  \begin{equation}
    \label{eq:manycooks-upper}
    R \leq \min_{{\cal S}\subset\{C_1,\ldots,C_{m-2}\}} S(A{\cal S}).
  \end{equation}
  This bound is an equality for all $m$, which is therefore
  the expression for the asymptotic version of the localisable entanglement.
  We prove this here for $m=4$; the general proof is given in~\cite{HOW}.
  Furthermore, the bound is achieved by a protocol
  where each helper $C_i$ takes a single turn in which he measures his state
  and communicate the result to the remaining parties.
\end{theorem}

Observe that the right hand side in eq.~(\ref{eq:manycooks-upper})
is the minimum pure state entanglement over all bipartite
cuts of the systems which separate Alice and Bob.

\medskip
The remainder of the paper is structured as follows: in
section~\ref{sec:EoA:protocol} we present a protocol and prove
theorem~\ref{thm:EoA:infty}. Section~\ref{sec:channel+helper}
presents an application of this first result to quantum
transmission with a classical helper in the channel environment.
In section~\ref{sec:EPR+GHZ:protocol} we show how to make the
basic protocol coherent, such that it also gives GHZ-states, and
prove theorem~\ref{thm:EPR+GHZ}. Its GHZ-rate we prove to be
optimal under a subclass of protocols which we call \emph{one-way
broadcast}. Then, in section~\ref{sec:manycooks} we generalise the
basic protocol to more than one helper (proof of
theorem~\ref{thm:n-party:EoA} for $m=4$), and discuss the connection
between the concept of localisable entanglement and entropy of
blocks of spins. Finally, in section~\ref{sec:normalform} we
discuss possible applications and/or extensions of our main result
to asymptotic normal forms of quantum channels, and conclude in
section~\ref{sec:discussion}.

\section{Asymptotic entanglement\protect\\ of assistance}
\label{sec:EoA:protocol}
In~\cite{E-of-A,cohen:EoA}, the following quantity was introduced
under the name of \emph{entanglement of assistance} of a bipartite
mixed state $\rho^{AB}$ (with purification $\psi^{ABC}$):
\begin{equation*}\begin{split}
  E_A(\rho^{AB})
      &:= E_A(\psi^{ABC})                                                                 \\
      &:= \max\left\{ \sum_i p_i E(\psi_i^{AB}) : \rho^{AB}=\sum_i p_i\psi_i^{AB} \right\}.
\end{split}\end{equation*}
The idea is that by varying a measurement (POVM) on $C$, the helper
Charlie can effect any pure state ensemble decomposition
$\rho^{AB} = \sum_i p_i\psi_i^{AB}$ for Alice and Bob's
state~\cite{schroedinger}.
In this sense, $E_A$ gives the maximum amount of entanglement
obtainable between Alice and Bob with the (remote) help from
Charlie. 
Of course, we are primarily interested in the operational
asymptotic rate of EPR states, $E_A^\infty$, which will
turn out to be given by the regularisation of $E_A$:
$$E_A^\infty(\rho) = \lim_{n\rightarrow\infty}
                          \frac{1}{n}E_A\bigl(\rho^{\otimes n}\bigr).$$

\medskip
Now we argue the upper bound, $E_A^\infty(\rho^{AB}) \leq S(A)$, which
was noted in~\cite{E-of-A}, operationally: whatever can be done under
three-party LOCC, is contained in protocols which allow general
transformations on $BC$ and LOCC with respect to the cut $A$ vs.~$BC$.
But in this latter formulation, we are in a bipartite pure state
situation, for which the maximum yield of EPR states is well-known
to be $S(A)$~\cite{BBPS}.
By an identical argument, we have the same bound with $S(B)$,
and hence we obtain
\begin{equation}
  \label{eq:entropy-upper}
  E_A^\infty(\rho^{AB}) \leq \min\bigl\{ S(A),S(B) \bigr\}.
\end{equation}

\medskip
Note that this upper bound is not additive under general tensor
products (compare~\cite{E-of-A}): consider strictly mixed states
$\rho^{AB}=\proj{0}^A\otimes\rho^B$ and
$\sigma^{A'B'}=\sigma^{A'}\otimes\proj{0}^{B'}$; they have both
$E_A=0$, because the entropy upper bound is $0$. However,
$E_A(\rho\otimes\sigma) > 0$. A less trivial example of
superadditivity of $E_A$ is given in~\cite{E-of-A} for two copies of
the same state. Here is a very easy one:

\begin{example}[Superadditivity of $E_A$]
  \label{ex:3-by-3}
  Consider the 3-qutrit determinant (or Aharonov) state
  $$\ket{\alpha}^{ABC} = \frac{1}{\sqrt{6}}
                            \bigl( \ket{012}+\ket{120}+\ket{201}
                                   -\ket{210}-\ket{102}-\ket{021} \bigr).$$
  The restriction $\alpha^{AB}$ is proportional to the projector
  onto the $3\times 3$-antisymmetric subspace, and it is well known
  that this subspace consists entirely of ``singlets'', i.e., states
  $\ket{v}\ket{v'}-\ket{v'}\ket{v}$, with $\langle v\ket{v'}=0$.
  Hence $E_A(\alpha) = 1$ (and by the way also
  $E_F(\alpha^{AB})=E_C(\alpha^{AB})=1$). However,
  $E_A(\alpha\otimes\alpha) \geq 2.5$, since $\alpha\otimes\alpha$
  can be presented as a uniform mixture of states
  $(U_1^{A_1}\otimes U_2^{A_2}\otimes U_1^{B_1}\otimes U_2^{B_2})\ket{\varphi}$, with
  \begin{align*}
    \ket{\varphi}^{A_1A_2B_1B_2}\!\!
           &=\frac{1}{\sqrt{8}}
               \bigl( (\ket{01} \!-\! \ket{10})^{A_1B_1}\otimes
                      (\ket{01} \!-\! \ket{10})^{A_2B_2} \bigr. \\
           &\phantom{===:}
               \bigl. + (\ket{12} \!-\! \ket{21})^{A_1B_1}\otimes
                        (\ket{12} \!-\! \ket{21})^{A_2B_2} \bigr) \! .
  \end{align*}
  It is easily established that the Schmidt spectrum of this state is
  $\left[ \frac{1}{4},\frac{1}{4},\frac{1}{8},\frac{1}{8},\frac{1}{8},\frac{1}{8} \right]$,
  so its entropy of entanglement is $E(\varphi) = 2.5$.
\end{example}

This example contains a valuable insight: for a given single-copy
decomposition of $\rho$, one can form superpositions of tensor
products of component states, and increase the entanglement.
A little consideration reveals that this is so because the
tensor products have some local distinguishability.
Hence, in the general case we should try to enforce local
distinguishability of the states we put in superposition.

\medskip\noindent
\begin{beweis}[of theorem~\ref{thm:EoA:infty}]
  Write $\ket{\psi}^{ABC} = \sum_j \sqrt{q_j}\ket{\psi_j}^{AB}\ket{j}^C$,
  with an orthogonal basis $\{\ket{j}\}$ of $C$. Let
  \begin{eqnarray}
  \label{eq:chi-A}
    \chi_A &= \chi\bigl\{ (q_j,\psi_j^A) \bigr\} = S(A) - \sum_j q_j S(\psi_j^A), \\
    \label{eq:chi-B}
    \chi_B &= \chi\bigl\{ (q_j,\psi_j^B) \bigr\} = S(B) - \sum_j q_j S(\psi_j^B),
  \end{eqnarray}
  denote the Holevo information of the given ensembles;
  observe the common term
  $$\sum_j q_j S(\psi_j^A) = \overline{E} = \sum_j q_j S(\psi_j^B).$$
  We may assume without loss of generality that $S(A)\leq S(B)$,
  hence $\chi_A \leq \chi_B$.

  For $n$ copies of $\psi$, the sequences $J=j_1\ldots j_n$,
  and consequently the states $\ket{J}=\ket{j_1}\cdots\ket{j_n}$,
  fall into (polynomially many) \emph{type classes}: we say that $J$
  is of type $P$ (which is a probability distribution on the letters $j$)
  if $j$ occurs exactly $nP(j)$ times in $J$.
  This is relevant because the probability $q_J=q_{j_1}\cdots q_{j_n}$,
  the product of the letter probabilities,
  of a sequence is constant across a type class.
  We can write the state as
  $$\bigl( \ket{\psi}^{ABC} \bigr)^{\ox n}
             = \sum_J \sqrt{q_J}\ket{\psi_J}^{A^nB^n}\ket{J}^{C^n},$$
  where $\ket{\psi_J}=\ket{\psi_{j_1}}\cdots\ket{\psi_{j_n}}$
  and $A^n=A_1 A_2\ldots A_n$ are Alice's $n$ copies of system $A$, etc.
  The goal of Charlie's strategy will be to project this state down
  to a superposition of terms $\ket{\psi_J}^{A^nB^n}$ which are
  as orthogonal as possible on both Alice's and Bob's systems: because
  then Alice's (say) reduced state is roughly an orthogonal mixture
  of the states $\psi_J^{A^n}$, and we can easily calculate its entropy.

  More precisely, Charlie's measurement consists of two steps:
  first, a projection into the subspaces of constant type, say $P$:
  $$\Pi(P) := \Span\bigl\{ J : J\text{ is of type }P \bigr\}.$$
  Note that, for any $\eta>0$, with probability $1-\e$, $\| P-q \|_1 \leq \eta$,
  if only $n$ is sufficiently large
  (otherwise, abort). Here, $\|\cdot\|_1$ is the total variational
  distance (or $1$-norm distance) of probability distributions.
  By Fannes' inequality (stated below as lemma~\ref{lemma:fannes}), then
  $$S\left( \sum_j P(j)\psi_j^A \right) -\sum_j P(j)S\bigl( \psi_j^A \bigr)
                                                        \geq \chi_A - \delta.$$

  Second, for each such type $P$, letting
  $N = \left\lfloor 2^{n(\chi_A-2\d)} \right\rfloor$, define states
  depending on a set $\cJ=\{J^{(0)},\ldots,J^{(N-1)}\}$ of sequences
  of type $P$ and a number $\alpha=0,\ldots,N-1$:
  $$\ket{t_\cJ(\alpha)} = \frac{1}{\sqrt{N}} \sum_{\beta=0}^{N-1}
                                e^{2\pi i \alpha\beta/N} \ket{J^{(\beta)}}.$$
  Clearly, with an appropriate constant $c>0$, the collection
  $\left( \frac{c}{N}\proj{t_{\cJ}(\alpha)} \right)_{\cJ,\alpha}$
  forms a POVM on the type $P$ subspace, i.e., these operators sum
  up to $\Pi(P)$. This is the second (rank-$1$!) POVM of Charlie.

  By the Holevo-Schumacher-Westmoreland theorem, stated as lemma~\ref{lemma:HSW}
  below, the majority of the sets $\cJ$ are good codes for the
  classical-quantum channel $j\longmapsto \psi_j^A$, and simultaneously
  for $j\longmapsto \psi_j^B$. For such a $\cJ$, and all $\alpha$,
  consider the projected state $\ket{\vartheta}$ of $AB$
  (up to normalisation), dropping the superscript $n$
  from the registers:
  $$\ket{\vartheta} = \frac{1}{\sqrt{N}} \sum_{\beta=0}^{N-1}
                         e^{-2\pi i\alpha\beta/N}\ket{\psi_{J^{(\beta)}}}^{AB}.$$
  Because Alice and Bob have good decoders for $\beta$,
  i.e., POVMs $(D_\beta^A)_\beta$ and $(D_\beta^B)_\beta$, they can
  locally extract $\beta$ with high reliability.
  We can always think of these measurements as (local) isometries,
  for example for Alice
  $$V_A = \sum_\beta \sqrt{D_\beta^A}\ox\ket{\beta}^{B'},$$
  and a similar expression $V_B$ for Bob.
  This pair of unitaries takes the state $\ket{\vartheta}$ to
  \begin{equation*}\begin{split}
    \ket{\tilde\vartheta} &= (V_A\ox V_B)\ket{\vartheta}   \\
                          &= \sum_{\beta,\gamma} \Bigl(
                                     \sqrt{D_\beta^A}\ox\!\sqrt{D_\gamma^B}\,\ket{\vartheta}
                                                 \Bigr)^{AB}\ox\ket{\beta}^{B'}\ket{\gamma}^{C'}.
  \end{split}\end{equation*}
  Since we have (with high probability) a good code both for Alice's
  and Bob's channels, we expect that
  $$\ket{\tilde\vartheta}
    \approx \frac{1}{\sqrt{N}} \sum_{\beta=0}^{N-1}
              e^{-2\pi i\alpha\beta/N}\ket{\psi_{J^{(\beta)}}}^{A^nB^n}
                                      \ket{\beta}^{A'}\ket{\beta}^{B'},$$
  which indeed can be shown
  (see the proof of theorem~\ref{thm:EPR+GHZ} below).
  Here we need something only slightly weaker:

  When tracing over $BB'$, we can assume that Bob's POVM is actually performed
  (i.e., the register $B'$ observed); using the fact that both Alice's and Bob's
  POVMs have average error probability $\leq \e$, we get
  \begin{equation*}
    \left\| \tilde\vartheta^A
           - \frac{1}{N}\sum_\beta \sqrt{D_\beta^A}\psi_{J^{(\beta)}}^A\sqrt{D_\beta^A}
                                                                   \ox \proj{\beta}^{A'}
    \right\|_1                                        \leq 2\cdot 2\e,
  \end{equation*}
  with the trace norm $\|\cdot\|_1$ on (density) operators.
  Furthermore, by the gentle measurement lemma~\ref{lemma:gentlePOVM},
  stated below for convenience, this yields
  \begin{equation*}
    \left\| \tilde\vartheta^A
           - \frac{1}{N}\sum_\beta \psi_{J^{(\beta)}}^A \ox \proj{\beta}^{A'}
    \right\|_1                             \leq 4\e + \sqrt{8\e} \leq 7\sqrt{\e}.
  \end{equation*}
  Hence, for the entropy (choosing $\e$ and $\eta$ small enough),
  \begin{equation*}\begin{split}
    S\bigl( \vartheta^A \bigr) &= S\bigl( \tilde\vartheta^A \bigr)                     \\
                   &\geq S\left( \frac{1}{N}\sum_\beta
                                 \psi_{J^{(\beta)}}^A \otimes \proj{\beta}^{A'} \right)
                                                                                 - n\d \\
                   &=    \log N
                         + \frac{1}{N}\sum_\beta E\bigl( \psi_{J^{(\beta)}}^{AB} \bigr)
                                                                                  -n\d \\
                   &=    \log N + n\sum_j P(j)E(\psi_j^{AB})                      -n\d \\
                   &\geq n\bigl( S(A) - 4\d \bigr),
  \end{split}\end{equation*}
  where we have used the Fannes inequality, the fact that all $J^{(\beta)}$
  have the same type $P$, and Fannes inequality once more.
\end{beweis}

\begin{lemma}[Fannes inequality~\cite{fannes}]
  \label{lemma:fannes}
  For any two states $\rho$ and $\sigma$ on a $d$-dimensional Hilbert space:
  if $\| \rho-\sigma \|_1 \leq \epsilon$, then
  $| S(\rho)-S(\sigma) | \leq \eta(\epsilon)+K\epsilon\log d$,
  with $\eta(x)=-x\log x$ and a universal constant $K$.
  \qed
\end{lemma}

\begin{lemma}[HSW theorem~\cite{H:SW}]
  \label{lemma:HSW}
  For a classical-quantum channel $W:x\longmapsto W_x$ on the
  Hilbert space ${\cal H}$, and a probability distribution
  $P$, let $U^{(i)}$ be i.i.d.~uniformly random from the sequences of
  length $n$ of type $P$. Then for every $\epsilon,\delta>0$
  and sufficiently large $n$, if $\log N\leq n\bigl( \chi\{(P(x),W_x)\}-\delta\bigr)$,
  $$\Pr\left\{ {\cal C}=(U^{(i)})_{i=1}^N \text{ is }\epsilon{-good} \right\}
                                                               \geq 1-\epsilon.$$
  Here we call a collection of codewords \emph{$\epsilon$-good} if there exists a POVM
  $(D_i)_{i=1}^N$ on ${\cal H}^{\otimes n}$ such that
  \begin{equation*}
    \frac{1}{N}\sum_{i=1}^N \tr\bigl( W^n_{U^{(i)}} D_i \bigr) \geq 1-\epsilon.
  \end{equation*}
  (In this form, the theorem is proved in~\cite{devetak:winter:HASHING}.)
  \qed
\end{lemma}

\begin{lemma}[Gentle measurements~\cite{winter:qstrong}]
  \label{lemma:gentlePOVM}
  Let $\rho$ be a state (actually, $\rho\geq 0$ and $\tr\rho\leq 1$
  are enough) and $0\leq X\leq \1$, such that $\tr(\rho X) \geq 1-\e$.
  \par\noindent
  Then, $\bigl\| \rho - \sqrt{X}\rho\sqrt{X} \bigr\|_1 \leq \sqrt{8\e}$.
  \qed
\end{lemma}

\section{Channel capacity with classical helper in the environment}
\label{sec:channel+helper}
Gregoratti and Werner~\cite{lost:found} have considered the following channel
model with helper in the environment:
$$U: {\cal H}_A \longrightarrow {\cal H}_B\otimes{\cal H}_C,$$
described by an isometry from Alice's input system $A$ to
the combination of Bob's output system $B$ and the environment $C$.
Assume that the environment system may be measured and the
classical results of the observation be forwarded to Bob --- attempting
to help him in error correcting quantum information sent from Alice.

\begin{figure}[ht]
  \centering
  \includegraphics[width=7.5cm]{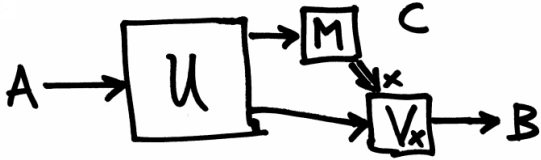}
  \caption{Alice prepares an input to (many copies of) the isometry
           $U$, which gives part of the state to Bob and part to Charlie.
           The latter measures a POVM $M$ on his system and classically
           communicates his result $x$ to Bob, who executes a unitary $V_x$
           depending on Charlie's message to recover Alice's sent state.}
  \label{fig:lost+found}
\end{figure}

We are interested in the quantum capacity of this scenario
from Alice to Bob, in the asymptotic limit of block coded
information (and collectively measured environment).
The setup is illustrated in figure~\ref{fig:lost+found}.

We want to mention a related model, discussed by Hayden and
King~\cite{hayden:king}, where the objective is to transmit classical
information rather than quantum. Of course, the corresponding
capacity will usually be higher, since the helper
in the environment can learn part
of the message and forward this information to Bob.

\begin{theorem}
  \label{thm:lost+found}
  The \emph{environment-assisted capacity} of a noisy quantum
  channel $T:A\longrightarrow B$ is
  $$C_A(T) = \max_\rho \, \min\bigl\{ S(\rho),S\bigl(T(\rho)\bigr) \bigr\}.$$
  The same capacity is obtained allowing unlimited LOCC between
  Alice, Bob and Charlie.
\end{theorem}
\begin{beweis}
  Let us first deal with the converse: whatever the detailed
  strategy, Alice will eventually input the $A^n$-part of
  some state $\ket{\Phi}^{A'A^n}$ into the channel (there is no loss of generality
  in assuming that the players keep all ancillas around, and hence the
  state pure). After the channel, the three players share the state
  $$\ket{\psi}^{A'B^nC^n} = \bigl( \1\ox U^{\ox n} \bigr)\ket{\Phi}^{A'A^n}.$$
  By the same argument as for the upper bound in theorem~\ref{thm:EoA:infty},
  the pure state entanglement between Alice and Bob cannot exceed either of
  \begin{align*}
    S\bigl( \psi^{A^n} \bigr) &\leq \sum_k S(\psi^{A_k})
                                  = \sum_k S(\Phi^{A_k})
                               \leq  n S(\rho^A)\ \text{ and}        \\
    S\bigl( \psi^{B^n} \bigr) &\leq \sum_k S(\psi^{B_k})
                                  = \sum_k S\bigl(T(\Phi^{A_k})\bigr)
                               \leq  n S\bigl( T(\rho^A) \bigr),
  \end{align*}
  with $\rho^A = \frac{1}{n}\sum_k \Phi^{A_k}$.

  For the direct part, let $\rho$ be the optimal
  input state for the maximum in the theorem and denote
  a purification of it $\ket{\phi}^{A'A}$, which is used in the
  following as ``test state''.

  Let Charlie pick, for some $n$, an optimal measurement
  $(M_x)_x$ for the entanglement of assistance of
  ($n$ copies of) $\ket{\psi} = (\1\ox U)\ket{\phi}^{A'A}$,
  according to theorem~\ref{thm:EoA:infty}.
  Then we can define a new quantum channel
  \begin{align*}
    T' : A^n     &\longrightarrow B^n B' \\
         \varphi &\longmapsto \sum_x \tr_{C^n}\!\bigl[ (M_x\ox\1)(U\varphi U^*) \bigr]
                                                                   \ox \proj{x}^{B'},
  \end{align*}
  which, by theorem~\ref{thm:EoA:infty}, has on the test
  state $\ket{\phi}^{\ox n}$ the
  \emph{coherent information}~\cite{schu:niel}
  \begin{equation*}\begin{split}
    I(A'\,\rangle B^n B')
      &= S(B^n B')-S(A' B^n B')    \\
      &\geq n \Bigl( \min\bigl\{ S(\rho), S\bigl(T(\rho)\bigr) \bigr\} - \d \Bigr).
  \end{split}\end{equation*}
  Invoking the quantum channel coding theorem~\cite{lloyd:Q,shor:Q,devetak:Q},
  there are block codes for $T'$ achieving this rate asymptotically.
\end{beweis}

\section{GHZ distillation}
\label{sec:EPR+GHZ:protocol}
Now we will show how to modify the protocol of theorem~\ref{thm:EoA:infty}
by ``making it coherent'' (after the model of~\cite{devetak:Q,devetak:winter:hashing})
such that part of its yield is in the form of GHZ states.
We shall freely use the notation introduced in the proof
of theorem~\ref{thm:EoA:infty}.

\medskip
\begin{beweis}[of theorem~\ref{thm:EPR+GHZ}]
  By possibly embedding $C$ into a larger space, we can write
  $\ket{\psi}^{ABC} = \sum_j \sqrt{q_j}\ket{\psi_j}^{AB}\ket{j}^C$,
  for any pure state decomposition of $\psi^{AB}$ into an
  ensemble $\{q_j,\psi_j^{AB}\}$.

  Consider sets $\cJ=\{J^{(0)},\ldots,J^{(N-1)}\}$ of
  $N = \left\lfloor 2^{n(\chi_A-2\d)} \right\rfloor$ type $P$
  sequences of length $n$, with, as before, $\| P-q \|_1 \leq \eta$.
  Now construct the projectors
  $$\Theta_{\cJ} = \sum_{\alpha=0}^{N-1} \proj{t_{\cJ}(\alpha)},$$
  so that we have a POVM $\bigl( c\,\Theta_{\cJ} \bigr)_{\cJ}$, a coarse-graining
  of the measurement used in the proof of theorem~\ref{thm:EoA:infty}.

  Charlie's measurement is again
  in two parts: first he measures the type subspace $\Pi(P)$, and
  $P$ is close to $q$ as above with high probability (otherwise
  abort). Then he measures $\bigl( c\,\Theta_{\cJ} \bigr)_{\cJ}$;
  if the operator $\Theta_{\cJ}$ acts, the projected state is
  (dropping the superscript $n$ from the register names)
  $$\ket{\zeta}^{ABC} = \frac{1}{\sqrt{N}} \sum_{\beta=0}^{N-1}
                           \ket{\psi_{J^{(\beta)}}}^{AB}\ket{J^{(\beta)}}^{C}.$$
  Most of the sets $\cJ$ are good codes for both the channels
  $j\longmapsto \psi_j^A,\,\, \psi_j^B$.
  Hence, with large probability, we can
  use the same local isometries $V_A$ and $V_B$ as before to
  extract $\beta$ with little state disturbance, and
  a local unitary $V_C$ mapping $\ket{J^{(\beta)}}\mapsto\ket{\beta}$.
  These isometries map $\ket{\zeta}$ to
  $$\ket{\tilde\zeta}^{AA'BB'C'} \approx
                                   \frac{1}{\sqrt{N}} \sum_{\beta=0}^{N-1}
                                       \ket{\psi_{J^{(\beta)}}}^{AB}
                                       \ket{\beta}^{A'}\ket{\beta}^{B'}\ket{\beta}^{C'},$$
  and because the $J^{(\beta)}$ are all of the same type, they
  are permutations of each other, so the states $\ket{\psi_{J^{(\beta)}}}^{AB}$
  can be taken to a standard state $\ket{\psi_{\overline{J}}}^{AB}$
  --- say, the lexicographically first sequence of type $P$ --- by
  (controlled) permutations of the $n$ subsystems.
  So, they arrive at the state
  $$\ket{Z}^{AA'BB'C'} \approx \ket{\psi_{\overline{J}}}^{AB}
                               \otimes \frac{1}{\sqrt{N}} \sum_{\beta=0}^{N-1}
                                         \ket{\beta}^{A'}\ket{\beta}^{B'}\ket{\beta}^{C'}.$$
  This concludes the proof, since the rate of $N$ is
  asymptotically $\chi_A$, and the rate of $E\bigl( \psi_{\overline{J}}^{AB} \bigr)$
  is asymptotically $\overline{E}$.
\end{beweis}

\begin{remark}
  \label{rem:communication}
  We have presented the POVMs of theorem~\ref{thm:EoA:infty}
  and~\ref{thm:EPR+GHZ} in the simplest possible terms. One can also
  minimise these POVMs, by not taking all sets $\cJ$. This can
  be done as shown in~\cite{devetak:winter:CR,devetak:winter:HASHING},
  yielding for theorem~\ref{thm:EoA:infty} a rank-$1$ measurement
  with $\approx 2^{n S(C)}$ elements; for theorem~\ref{thm:EPR+GHZ}
  the POVM has $\approx 2^{n[H(q)-\chi_A]}$ operators.
\end{remark}

Now we show that theorem~\ref{thm:EPR+GHZ} is in a certain sense
optimal: namely, it gives the largest GHZ rate among all protocols
which consist only of (i) a local operation with measurement at
$C$, (ii) sending the classical information obtained in the measurement
to $A$ and $B$, and (iii) local operations of $A$ and of $B$
depending on the message. In particular, we allow no feedback
communication and no communication between Alice and Bob.
These are severe restrictions, but at least the protocol from
theorem~\ref{thm:EPR+GHZ} is of this type: we call
it \emph{one-way broadcast}.

\begin{theorem}
  \label{thm:oneway-bc}
  Under one-way broadcast protocols from $C$ to $AB$,
  the asymptotic GHZ rate from the state $\psi^{ABC}$ cannot
  exceed $\min\{ S(A),S(B) \} - E_C\bigl(\psi^{AB}\bigr)$.
\end{theorem}
\begin{beweis}
  We show actually a bit more: the rate of three-way
  \emph{common randomness} distillable by such protocols is
  asymptotically bounded by the same number.
  This problem was studied in~\cite{devetak:winter:CR} for
  two players with one-way communication, and the relevant observation
  here is that with one-way broadcast, the task is equivalent to
  two simultaneous two-player common randomness distillations:
  from $C$ to $A$ and from $C$ to $B$.

  What it is about is the following: the sender $C$ and the receiver
  ($A$ or $B$)
  initially share a quantum state, and by local operations and
  one-way classical communication want to distill a maximum
  amount of shared randomness, which however has to be independent
  of the communicated message(s).

  A particular protocol for doing this is to distill GHZ states
  by a one-way broadcast protocol and then all three measure
  these states in the computational basis --- by purity of
  the measured state, the resulting perfect shared randomness
  is independent of everything else in the protocol.

  It was shown in~\cite{devetak:winter:CR}
  that the maximum rate achievable between $C$ and $A$ is the
  maximum of eq.~(\ref{eq:chi-A}) --- actually regularised for
  many copies of the state; and similarly between $C$ and $B$
  the --- regularised --- maximum of eq.~(\ref{eq:chi-B}).
  The smaller of these numbers clearly is just
  $\min\{ S(A),S(B) \} - E_C\bigl(\psi^{AB}\bigr)$.
\end{beweis}

\begin{remark}
  \label{rem:separable}
  In general, the GHZ rate obtainable from a pure state $\ket{\psi}^{ABC}$ by
  general LOCC has the easy upper bound $\min\bigl\{ S(A),S(B),S(C) \bigr\}$.
  Remarkably, our protocol of theorem~\ref{thm:EPR+GHZ} achieves this
  in for broad class of states, namely when one of the reduced states
  $\psi^{AB}$, $\psi^{BC}$ or $\psi^{AC}$ is separable.
\end{remark}

\begin{example}[Groisman, Linden, Popescu~\cite{bristolians}]
  \label{expl:interpolating}
  Consider the family of states
  $$\ket{\Upsilon_\alpha}^{ABC}
         = \alpha\ket{0}^A\ket{\Phi^+}^{BC}+\beta\ket{1}^A\ket{\Phi^-}^{BC},$$
  with $0\leq\alpha\leq\beta$ and $\alpha^2+\beta^2=1$, interpolating between
  a state $\ket{\EPR}^{BC}$ ($\alpha^2=0$) and $\ket{\GHZ}^{ABC}$
  ($\alpha^2=1/2$; up to local unitaries).
  Observe that it is certainly possible to
  obtain $1$ EPR state between $B$ and $C$ from $\Upsilon_\alpha$.

  In~\cite{bristolians} it is observed that the local entropies of
  $\Upsilon_\alpha$ are consistent with the hypothetical existence of an
  asymptotically reversible transformation into
  \begin{equation}
    \label{eq:conjecture}
    H_2(\alpha^2)                      \, \times \ket{\GHZ}^{ABC}\text{ and }
    \bigl[ 1 \!-\! H_2(\alpha^2) \bigr]\, \times \ket{\EPR}^{C}
  \end{equation}
  per copy of the state,
  but the authors present heuristic arguments for its impossibility.

  Let us see what our results tell us about the distillability of
  GHZ and EPR states: by applying theorem~\ref{thm:EPR+GHZ} with
  $B$ (or equivalently $C$)
  in the role of the helper, we obtain (since $\Upsilon^{AC}$ is
  separable) a GHZ rate of $H_2(\alpha^2)$, but no EPR states.
  The GHZ rate is evidently optimal under general LOCC protocols,
  as it coincides with Alice's
  entropy (i.e., her entanglement with the rest of the players).
  By applying theorem~\ref{thm:EPR+GHZ} with $A$ as the helper,
  we have to calculate the entanglement cost of the Bell
  mixture $\Upsilon^{BC}$,
  which happens to be known by~\cite{VDC:additive} and ~\cite{Wootters:EoF}:
  $$E_C\bigl(\Upsilon^{BC}\bigr) = E_F\bigl(\Upsilon^{BC}\bigr)
                                 = H_2\!\left( \frac{1}{2}-\alpha\beta \right).$$
  Hence we get distillation of
  $$\left[1 \!-\! H_2\!\left(\frac{1}{2}-\alpha\beta\right)\right]
                                             \, \times \ket{\GHZ}^{ABC}\text{ and }
    H_2\!\left(\frac{1}{2}-\alpha\beta\right)\, \times \ket{\EPR}^{C}$$
  per copy of the state, and theorem~\ref{thm:oneway-bc}
  shows that this GHZ rate is optimal
  among all one-way broadcast protocols from $A$ to $BC$.
  Note that the GHZ rate is slightly worse than the one
  stated in eq.~(\ref{eq:conjecture}), of which it is unknown if
  it can be achieved.
\end{example}

\begin{example}[$W$-State]
  \label{expl:W}
  Another interesting example is provided by the W-state~\cite{acin:etal}
  $$\ket{W} = \frac{1}{\sqrt{3}}\bigl( \ket{001}+\ket{010}+\ket{100} \bigr),$$
  which is interesting because it cannot be converted to a GHZ state
  even probabilistically (on a single copy).

  Theorem~\ref{thm:EoA:infty} tells us that any two parties can obtain
  a rate of $H_2\bigl(\frac{1}{3}\bigr) \approx 0.918$
  EPR states, with assistance from the third.
  Since two EPR pairs between different players can be converted
  into a GHZ state, we can obtain a GHZ rate of at least
  $\frac{1}{2}H_2\bigl(\frac{1}{3}\bigr) \approx 0.459$.

  However, using theorem~\ref{thm:EPR+GHZ}, we can do a bit better:
  the entanglement of formation of any two-party reduced state is
  evaluated with the help of~\cite{Wootters:EoF}, and we get rates of
  $H_2\bigl(\frac{1}{3}\bigr) - H_2\bigl( (1-\sqrt{5/9})/2 \bigr) \approx 0.368$
  for GHZ states,
  and $H_2\bigl( (1-\sqrt{5/9})/2 \bigr) \approx 0.550$ for EPR states
  between any pair of players.
  Converting the EPR states to GHZ's as before, we arrive at an overall
  rate of GHZ states of
  $H_2\bigl(\frac{1}{3}\bigr) - \frac{1}{2} H_2\bigl( (1-\sqrt{5/9})/2 \bigr)$,
  which is $\approx 0.643$.
\end{example}

\section{Singlet distillation with the help of many (distant) friends;\protect\\
         asymptotic localisable entanglement}
\label{sec:manycooks}
Consider now the $m$-party generalisation of the task 1: distillation
of EPR pairs between $A$ and $B$ with the help of $C_1,\ldots C_{m-2}$
from an $m$-partite pure state, by LOCC.
\par
In analogy to the upper bound eq.~(\ref{eq:entropy-upper}),
we can easily obtain an upper bound
on the achievable rate $R$ in this scenario: surely, the distillable entanglement
can only go up if we allow Alice to team up with a subset ${\cal S}$ of the
helpers $C_i$, and Bob with the complement
$\overline{\cal S} = \{1,\ldots,m-2\}\setminus{\cal S}$, such that
all collective operations on $A{\cal S}$ and on $B\overline{\cal S}$ are
allowed, and LOCC between these two groups.
Thus, $R\leq S(A{\cal S})$, and we get eq.~(\ref{eq:manycooks-upper}),
$$R \leq \min_{{\cal S}} S(A{\cal S}) = \min_{{\cal S}} S(B\overline{{\cal S}}).$$
[Note that this reduces to the inequality~(\ref{eq:entropy-upper}) for $m=3$.]

In~\cite{Ashish:JSmo} it was shown that whenever the right hand side
in the above equation is non-zero, then one Alice and Bob can, with
LOCC help from the other parties, distill EPR pairs at non-zero rate.

It turns out, however, that the right hand side is achievable for any $m$,
and we show here how to do it in the case of $m=4$ (the general case
requires different arguments and is solved in~\cite{HOW}).

\medskip
\begin{beweis}[of theorem~\ref{thm:n-party:EoA} for $m=4$]
  Only the achievability of the minimum cut entanglement is left to
  be proved. For $m=4$, i.e.~two helpers Charlie and Debbie, this means
  we are looking at $R = \min\{ S(A),S(B),S(AC),S(BC) \}$.

  Our goal will be to construct a measurement on $D^n$ such that
  for the (majority of) projected states $\ket{\vartheta}^{A^nB^nC^n}$,
  $$\min\left\{ S\bigl(\vartheta^{A^n}\bigr), S\bigl(\vartheta^{B^n}\bigr) \right\}
                                                                       \geq n(R-\d),$$
  with arbitrary $\d>0$ and sufficiently large $n$. I.e., we want
  to preserve (up to a small loss) the minimum cut entanglement,
  while disengaging Debbie. If we succeed doing this, we can invoke
  theorem~\ref{thm:EoA:infty} for the residual tripartite state.

  Pick any basis of $D$, so that we can write
  $\ket{\psi}^{ABCD} = \sum_j \sqrt{q_j}\ket{\psi_j}^{ABC}\ket{j}^D$.
  As in the previous proofs, we have reduced state ensembles
  with Holevo informations $\chi_A$, $\chi_B$, $\chi_{AC}$
  and $\chi_{BC}$. By possibly swapping $A$ and $B$, we may assume
  that $\chi_A \leq \chi_B$.
  Invoking monotonicity of the Holevo information under partial trace,
  $\chi_A \leq \chi_{AC}$ and $\chi_B \leq \chi_{BC}$, we are left
  with one of the following orderings of the four quantities:
  \begin{align*}
                 &\chi_A \leq \chi_B \leq \chi_{AC} \leq \chi_{BC}, \\
    \text{or }\  &\chi_A \leq \chi_{AC} \leq \chi_B \leq \chi_{BC}.
  \end{align*}

  Define $\chi_0 := \min\{ \chi_B,\chi_{AC} \}$ and consider random
  codes $\cJ$ of rate $\chi_0-\d$ where, in slight variation
  to the proof of theorem~\ref{thm:EoA:infty}, the codewords are drawn
  from the distribution $q^{\ox n}$ --- this is the original
  form of the HSW theorem~\cite{H:SW}, and the conclusion of lemma~\ref{lemma:HSW}
  holds true.
  Now construct a rank-$1$ measurement on $D$, in the same
  way as we did there.
  What can we say about the projected state
  $$\ket{\vartheta}^{ABC} = \frac{1}{\sqrt{N}}\sum_\beta
                              e^{-2\pi i\alpha\beta/N}\ket{\psi_{J^{(\beta)}}}^{ABC}\ ?$$

  For the bipartition $B|AC$, essentially the same
  argument from that proof shows that
  with high probability, the entanglement of $\vartheta$ is
  $$E(\vartheta^{B|AC}) \geq n\bigl( \min\{S(B),S(AC) \} -\d \bigr)
                                                       \geq n(R-\d).$$

  For the bipartition $A|BC$ this only works when $\chi_A=\chi_0$,
  to make the rate of the codes smaller than either Holevo
  information. So, let us assume $\chi_A < \chi_0$, and $\d$
  so small that $\chi_A+\d \leq \chi_0-\d$. The rate of the code is
  still smaller than $\chi_{BC}$, so there exists a (hypothetical)
  ``local'' decoding of $\beta$ from the register $BC$.
  I.e., with respect to the bipartite cut $A|BC$, the
  state $\ket{\vartheta}$ is equivalent to
  $$\ket{\tilde\vartheta}^{ABC} \approx \frac{1}{\sqrt{N}}\sum_\beta
                                    e^{-2\pi i\alpha\beta/N}\ket{\psi_{J^{(\beta)}}}^{ABC}
                                                            \ket{\beta}^{\widetilde{BC}},$$
  where the approximation has the same quality as in the proof of
  theorem~\ref{thm:EoA:infty}. But then, we have
  $$\tilde\vartheta^A \approx \frac{1}{N}\sum_\beta  \psi_{J^{(\beta)}}.$$

  Now we can conclude the proof by invoking lemma~\ref{lemma:opChernoff}
  below, which states that for a random code (which is what the POVM
  will select) the average on the right hand side is
  $\approx \bigl( \psi^A \bigr)^{\ox n}$. Hence, and using Fannes' inequality
  once more,
  $$E(\vartheta^{A|BC}) \geq n\bigl( S(A)-\d \bigr) \geq n(R-\d),$$
  and we are done.
\end{beweis}

\begin{lemma}[Density sampling~\cite{A-W:ID}]
  \label{lemma:opChernoff}
  Consider the ensemble $\bigl\{ (q_j,\rho_j) \bigr\}$ of states
  on a $d$-dimensional Hilbert space, with average
  density operator $\rho$ and Holevo information $\chi$.
  Let independent and identically
  distributed random variables $X_1,\ldots,X_N$,
  drawn from the states
  $\rho_J = \rho_{j_1}\ox\cdots\ox\rho_{j_n}$ with probability
  $q_J = q_{j_1}\cdots q_{j_n}$.
  Then, for every $\e,\d>0$, $N \geq 2^{n(\chi+\d)}$, and
  sufficiently large $n$,
  \begin{equation*}
    \left\| \frac{1}{N}\sum_{k=1}^N X_k - \rho^{\ox n} \right\|_1 \leq \e
  \end{equation*}
  with probability $\geq 1-\e$
  \qed
\end{lemma}

Theorem~\ref{thm:n-party:EoA} yields an exact expression for the asymptotic
localisable entanglement~\cite{localiz} (except for the technical
issue that there one has an infinite number of parties, whereas
here we considered only finite $m$). The concept of
localisable entanglement was introduced in the context of quantum
spin systems, and allows to define a notion of entanglement length
when these spins are part of a lattice with a given geometry. More
precisely, consider the maximal bipartite entanglement that can be
localised between two blocks of spins as a function of the
distance between the blocks; typically this function is decaying
exponentially with the distance, $\exp(-L/\xi)$, and the
entanglement length is defined as the constant $\xi$ in this
exponent. Theorem~\ref{thm:n-party:EoA} gives the exact expression for the localisable
entanglement between two blocks if asymptotically many
realisations of these systems are available and \emph{joint local}
operations can be performed. If furthermore we are considering a
state with infinitely many particles and translational symmetry
(which is the usual case in condensed matter systems), then the
strong subadditivity property of the von Neumann entropy enforces
the entropy of a block of spins to grow when more spins are
included in the block. It follows that the asymptotic localisable
entanglement in such systems between two blocks is exactly given
by the minimal entropy of these blocks, which proves that the
upper bound given in~\cite{localiz} is actually the exact value
for the localisable entanglement in the asymptotic limit. This is
very surprising: the magic of doing local asymptotic operations
allows to \emph{create} entanglement between two blocks that are
arbitrary far from one another, and the rate at which this can be
done is independent of the distance. This implies that any
nontrivial ground state can be used as a perfect quantum repeater
if many copies are available in parallel. The amount of
entanglement that can be localised over these arbitrary distance
is solely related to the entropy of a block of spins and not
dependent on the distance. It is interesting to contrast the
translationally invariant case to the one with random bond
interactions~\cite{MooreandRafael}; in the latter case, the
minimal entropy over all bipartite cuts will decrease
algebraically with the distance between the blocks. This indicates
that the entanglement in the case of random systems is essentially
different than in the case of translationally invariant ones,
something that is not revealed by looking at the entropy of a
block of spins.

The problem of calculating the entropy of a block of spins has
recently attracted a lot of attention in condensed
matter physics~\cite{Vidalandco}, where it was shown that this
entropy, in the case of ground states of 1-dimensional systems,
saturates to a finite value or increases logarithmically as a
function of the size of the block, depending on whether
the system is critical or not. The present work provides an
operational meaning to these calculations in the sense of
entanglement theory: this entropy quantifies the amount of
entanglement that can be created at arbitrary distances if this
ground state would be used as a quantum repeater. In higher
dimensional systems, the entropy of a block of spins grows as the
boundary of that block, and therefore there is no bound on the
amount of EPR-pairs that could be localised between two far away
regions by doing \emph{joint local} measurements on all the other
spins; this is again the consequence of the fact that the
asymptotic operations allow for perfect entanglement swapping in
multipartite states.

\section{Asymptotic normal forms of unital \&{} general quantum channels}
\label{sec:normalform}
Based on the well-known linear isomorphism
between completely positive and trace preserving maps and a set of
quantum states~\cite{jamiolkowski}:
$$T:A\rar B \quad \Longleftrightarrow \quad \rho^{A'B} = \rho_T = (\id\ox T)\phi^{A'A},$$
with a pure state $\ket{\phi}^{A'A}$ of Schmidt-rank
$d_A = \dim{\cal H}_A$, we can interpret our findings in
theorem~\ref{thm:EoA:infty} as statements on quantum channels.
Note that $T$ is an isometry or unitary, if and only if
the state $\rho_T$ is pure, and the corresponding states
of different isometries are equivalent to each other
up to unitaries on the system $B$. We shall use this isomorphism
in the following with a maximally entangled state $\phi^{A'A}$
of Schmidt rank $d_A$, unless specified otherwise.

Let $T$ be a unital quantum channel on a system, i.e. mapping the
identity on $A$ to the identity on $B$ (and assume
input and output system to be of the same dimension $d=d_A=d_B$
for the moment). The corresponding state has the properties
$\rho^{B} =  \tr_{A'} \rho_T = \frac{1}{d}\1$,
$\rho^{A'} = \tr_B    \rho_T = \frac{1}{d}\1$.
Thanks to theorem~\ref{thm:EoA:infty}, we know that in the asymptotic scenario
the entanglement of assistance of $\rho_T$ is given by $\log d$.

Clearly, the unital channels (and equally, the states with maximally
mixed marginals as above), for a convex set, and the question
of determining its extremal points has attracted quite some
attention~\cite{Streater}. The classical analogue of this problem
is about doubly stochastic maps which, thanks to Birkhoff's theorem,
are known to be exactly the convex combinations of permutations.
For quantum doubly stochastic maps (another popular name for
unital trace-preserving channels) the ``obvious'' generalisation is wrong:
there exist unital channels which are \emph{not} convex combinations
of unitaries. Under the Jamio\l{}kowski isomorphism, this means that
the state $\rho_T$ is not a convex combination of maximally
entangled states; a specimen of this type we have actually studied in
example~\ref{ex:3-by-3}.

However, theorem~\ref{thm:EoA:infty} points a way to resolving this
unsatisfactory state of affairs in the asymptotic limit: since the
asymptotic entanglement of assistance of $\rho_T$ is $\log d$,
we can say that $\rho_T^{\ox n}$ is well approximated by a convex
combination of ``almost'' maximally entangled states in the sense
that their entropies of entanglement are $n(\log d - \delta)$
for arbitrarily small $\delta>0$ and sufficiently large $n$.
We would like to deduce from this that $T^{\ox n}$ is well
approximated by a convex combination of unitaries (in the appropriate
norm), but unfortunately the latter is really a stronger statement
since it would give an approximation of $\rho_T^{\ox n}$ by a convex
combination of states that have high fidelity to some maximally entangled
state. And that is not even mentioning the issues of the different
norms to be used for comparing states and for channels.

Similarly, for a general channel, and general $\phi^{A'A}$,
the state $\rho_T^{\ox n}$ can be restricted to the typical
subspaces~\cite{quantum:coding} of $\bigl( \rho_T^{A'} \bigr)^{\ox n}$
on Alice's side, and of $\bigl( \rho_T^B \bigr)^{\ox n}$ on Bob's
side, without changing the state very
much. This projected state lives in a $D_A\times D_B$-dimensional
system, with $D_A \approx 2^{nS(A)}$ and $D_B \approx 2^{nS(B)}$.
By theorem~\ref{thm:EoA:infty} it is well approximated by
a convex combination of pure states with entanglement
$n\bigl( \min\{S(A),S(B)\} - \delta \bigr)$, which again
is too weak to say that the components have high fidelity
with maximally entangled states.

Nevertheless, we take these observations as positive evidence
for the following conjecture:

\begin{conjecture}
  \label{conj:unital}
  Let $T$ be a unital quantum channel on a system, or
  more generally a map $T:A\rar B$ such that for all
  input states $\rho^A$, $S(\rho) \leq S\bigl( T(\rho) \bigr)$.
  Then, for sufficiently large $n$, $T^{\ox n}$ is arbitrarily
  well approximated by mixtures of isometries (unitaries in
  the unital case).

  In general, $T^{\ox n}$ is arbitrarily well approximated by
  mixtures of partial isometries between $A^n$ and $B^n$
  (i.e., unitary transformations between subspaces of systems $A^n$
  and $B^n$).

  The appropriate distance measure for quantum channels $T$ and $T'$
  to be used here is
  $$\| T - T' \|_{\rm cb} = \max_{\phi} \| (\id\ox T)\phi - (\id\ox T')\phi \|_1,$$
  the completely bounded norm (cb-norm)~\cite{cb-norm}.
\end{conjecture}

In further support of this conjecture, we now outline
a proof for a weaker version of it, where the comparison of
$T^{\ox n}$ and the mixture $T'$ of unitaries is done not in the worst
case over all input states, but with respect to a single
state $\phi^{\ox n}$: we want
$\bigl\| \rho_T^{\ox n} - (\id\ox T')\phi^{\ox n} \bigr\|_1 \leq \epsilon$.
The significance of such a statement is that if $\phi$ is a
purification of a mixed state $\sigma$ on $A$, and
$\{ p_k,\phi_k \}$ is any source ensemble on $A^n$ with average
$\sigma^{\ox n}$, then the average error,
$\sum_k p_k \bigl\| T^{\ox n}(\phi_k) - T'(\phi_k) \bigr\|_1$,
is also bounded by $\epsilon$.

For simplicity, we assume the $S(\sigma)$ is strictly smaller than
$S\bigl(T(\sigma)\bigr)$.
Note that one could always modify the channel trivially
by padding the output with a sufficiently maximally mixed state,
to enforce this condition.

We can write down a purification of $\rho_T$ in Schmidt form,
$$\ket{\psi}^{A'BC} = \sum_j \sqrt{q_j} \ket{\psi_j}^{A'B}\ket{j}^C,$$
with orthogonal states $\{\ket{j}\}_j$ and $\{\ket{\psi_j}\}_j$.
By assumption,
\begin{equation*}\begin{split}
  \chi_{A'} &:= S(\sigma)
                     - \sum_j q_j S\bigl( \psi_j^{A'} \bigr) \\
            &<  S\bigl( T(\sigma) \bigr)
                     - \sum_j q_j S\bigl( \psi_j^{B} \bigr) =: \chi_B,
\end{split}\end{equation*}
so we can choose a number $R$ between these two values.
Now we go through the random coding argument in the proof
of theorem~\ref{thm:EoA:infty}, but actually in the form of
the second case considered in the proof of theorem~\ref{thm:n-party:EoA}.
Since here we assume that we have uniform distribution on the
$j$, there is no need to restrict to the set of typical sequences.

What we get are random codes of $N=2^{nR}$ sequences $J=j_1\ldots j_n$,
such that the corresponding states $\psi_{J^{(\beta)}}^B$
(dropping superscript $n$ as before) form a good code for Bob.
That means, that for the superpositions
$$\ket{\vartheta}^{A'B} = \frac{1}{\sqrt{N}}
                          \sum_\beta e^{-2\pi i \alpha\beta} \ket{\psi_{J^{(\beta)}}}^B$$
(resulting from projecting the system $C$ onto a vector $\ket{t_\vartheta}$),
we obtain, as at the end of the proof of theorem~\ref{thm:n-party:EoA},
that $\vartheta^{A'} \approx \frac{1}{N} \sum_\beta \psi_{J^{(\beta)}}^{A'}$.
And exactly as there, we can use lemma~\ref{lemma:opChernoff} to conclude
that $\vartheta^{A'} \approx \sigma^{\ox n}$ with respect to trace
distance.
Both approximations in fact with high probability over the
choice of the code.
That means that there is a purification $\ket{\zeta_\vartheta}^{A'B}$
of $\sigma^{\ox n}$ such that $\vartheta^{A'B} \approx \zeta_\vartheta^{A'B}$.

The connection to channels is now made by going the Jamio\l{}kowski
isomorphism in the other direction: for the well-behaved
$\vartheta$ as above, there exists an isometry $V_\vartheta:A\rar B$
such that $\ket{\zeta_\vartheta}=(\1\ox V_\vartheta)\ket{\phi^{\ox n}}$,
and hence our candidate mixture of unitaries is
$$T'(\varphi) = \sum_{\vartheta\text{ well-behaved}}
                         w_\vartheta\, V_\vartheta \varphi V_\vartheta^\dagger,$$
where the $w_\vartheta$ are probability weights. They are obtained
as essentially $\bigl| {}^C\!\bra{t_{\vartheta}} \psi^{\ox n}\rangle^{A'BC} \bigr|^2$,
normalised to the probability of the well-behaved set.
It is then straightforward to verify that indeed
$\bigl\| \rho_T^{\ox n} - (\id\ox T')\phi^{\ox n} \bigr\|_1$ is small.

\medskip
We want to close this section with a few comments on the difficulties
encountered in the attempt to extend this argument to a proof of
our conjecture. Clearly, the vectors $\ket{t_\vartheta}$ determine
Kraus operators $D_\vartheta$ for $T^{\ox n}$ via
$${}^C\!\bra{t_{\vartheta}} \psi^{\ox n}\rangle^{A'BC}
                    = (\1\ox D_\vartheta)\ket{\phi^{\ox n}}.$$
Because the cb-norm difference of $T^{\ox n}$ and $T'$ can be
upper bounded by $\sum_\vartheta \| D_\vartheta - V_\vartheta \|$
(with the operator norm $\|\cdot\|$),
it is tempting to aim at making the latter quantity small.
But with our fixed source $\sigma$, we can say something
about the difference $D_\vartheta - V_\vartheta$ only on
the typical subspace of the source, and even there only
on the average --- it is conceivable, and consistent with
our result, that the operator norms of the $D_\vartheta - V_\vartheta$,
when restricted to the typical subspace, are all large.
In addition, in the above proof we can make our statements
about $\vartheta$ only ``with high probability'' and the
probability distribution is also determined by the source
$\sigma$.

\section{Discussion}
\label{sec:discussion}
We have presented a class of very general procedures to distill
singlets and cat states, both in tripartite and multipartite
settings. These procedures give universally the largest EPR
rate distillable between any pair of parties in a multipartite state,
when the other players cooperate. For three parties, this
problem and its solution is equivalent to the previously
considered entanglement of assistance.
We have shown how GHZ (and higher cat state) distillation
protocols can be constructed from common randomness
distillation schemes by ``coherification''. It should be clear
that a good number of variations of what we have shown
here can be done.
As a consequence, we could solve the problem of quantum channel
coding with maximal classical help from the environment.

We stress that even though we look here at pure state transformations,
we did not attempt ``entanglement concentration'', which is meant
to generalise the asymptotic theory of bipartite pure states:
there we have asymptotic reversibility, and demanding this leads
to the hard MREGS problems. Instead, we do ``distillation''
of specific states (which surely will be interesting), embracing
the possibility of irreversibility, but going for the maximum
rate. We think that understanding these problems will remain
central even assuming availability of a complete MREGS.

Thus, starting from the strange entanglement of assistance
problem, we discovered a great number of highly interesting
results of multi-party entanglement processing. These also shed some
new light on issues like the entanglement length in spin chains.
Perhaps even more important are the conceptual insights regarding
possible asymptotic normal forms of quantum channels as mixtures
of partial isometries. Finally, we want to mention a spin-off in
quite another direction: based on the techniques of
section~\ref{sec:manycooks}, and developing them further, the
problem of \emph{distributed quantum data compression}
(with unlimited classical side communication) could be
solved in~\cite{HOW}. The methods of that paper also simplify some
of our arguments regarding EPR distillation (they don't apply to
GHZ distillation, however), and allow us to prove the equality in
theorem~\ref{thm:n-party:EoA} for all $m$.
\vspace{1cm}\\

\acknowledgments
We wish to thank Charles H. Bennett,
Ignacio Cirac, Igor Devetak, Mark Fannes, Micha\l{} Horodecki,
Debbie Leung, Jonathan Oppenheim and Tobias Osborne for interesting discussions
on entanglement of assistance and unital channels, and especially
Berry Groisman, Noah Linden and Sandu Popescu for sharing their
results in~\cite{bristolians} prior to publication.

JAS acknowledges the support of the NSA and ARO under contract number
DAAD19-01-C-0056.
AW is supported by the EU project RESQ (contract
no.~IST-2001-37559) and by the U.K. Engineering and Physical
Sciences Research Council's ``IRC QIP''.  The hospitality of the
Isaac Newton Institute of Mathematical Sciences, Cambridge, during
the topical semester on Quantum Information Sciences (16/08-17/12
2004) are gratefully acknowledged by JAS and AW. FV acknowledges
support by the Gordon and Betty Moore Foundation (the Information
Science and Technology Initiative, Caltech).

\end{document}